\pdfoutput=1
\documentclass[]{amsart}

\usepackage{lscape}
\usepackage{subcaption}
\usepackage{scalerel}
\usepackage{mathrsfs}
\usepackage{hyphenat}
\usepackage{braket}
\usepackage{mlmodern}
\usepackage{mathpartir}
\usepackage{doi}

\usepackage{tikz}
\usepackage{tikz-cd}
\usepackage{quiver}

\tikzset{every picture/.style={line width=0.85pt}}
\usepackage[leftmargin=1em,rightmargin=1ex,vskip=1ex]{quoting}
\usepackage[utf8]{inputenc}
\usepackage{cmll}

\newcommand\scalemath[3]{\scalebox{#1}[#2]{\mbox{\ensuremath{\displaystyle #3}}}}

\newcommand{\leftarrowtip}{\ensuremath{\tikz\draw[line width=0.5pt,->] (10pt,0) -- (0,0);}}
\newcommand{\leftarrowtailnotip}{\ensuremath{\tikz\draw[line width=0.5pt,-<] (0,0) -- (10pt,0);}}

\newcommand{\unicodeStar}{\ensuremath{\star}}
\DeclareUnicodeCharacter{2605}{\unicodeStar}

\DeclareUnicodeCharacter{21D2}{\ensuremath{\Rightarrow}}
\DeclareUnicodeCharacter{2218}{\ensuremath{\circ}}
\DeclareUnicodeCharacter{2248}{\ensuremath{\approx}}
\DeclareUnicodeCharacter{2022}{\ensuremath{\bullet}}
\DeclareUnicodeCharacter{2219}{\ensuremath{\bullet}}
\DeclareUnicodeCharacter{2026}{\ensuremath{\dots}}
\DeclareUnicodeCharacter{2208}{\ensuremath{\in}}
\DeclareUnicodeCharacter{2192}{\ensuremath{\to}}
\DeclareUnicodeCharacter{2190}{\ensuremath{\leftarrowtip}}
\DeclareUnicodeCharacter{2919}{\ensuremath{\leftarrowtailnotip}}

\DeclareUnicodeCharacter{00D7}{\ensuremath{\times}}
\DeclareUnicodeCharacter{00B7}{\ensuremath{\cdot}}
\DeclareUnicodeCharacter{222B}{\ensuremath{\int}}
\DeclareUnicodeCharacter{22A4}{\ensuremath{\top}}
\DeclareUnicodeCharacter{22A5}{\ensuremath{\bot}}
\DeclareUnicodeCharacter{2264}{\ensuremath{\leq}}

\newcommand{\unicodecolon}{\ensuremath{\colon}}
\DeclareUnicodeCharacter{FE55}{\unicodecolon}
\newcommand{\unicodeleftpar}{\ensuremath{\left(}}
\DeclareUnicodeCharacter{27EE}{\unicodeleftpar}
\newcommand{\unicoderightpar}{\ensuremath{\right)}}
\DeclareUnicodeCharacter{27EF}{\unicoderightpar}
\DeclareUnicodeCharacter{2260}{\neq}
\DeclareUnicodeCharacter{22A9}{\Vdash}
\DeclareUnicodeCharacter{2237}{\proportion}
\DeclareUnicodeCharacter{2124}{\mathbb{Z}}
\DeclareUnicodeCharacter{27E8}{\langle}
\DeclareUnicodeCharacter{27E9}{\rangle}
\DeclareUnicodeCharacter{21A6}{\mapsto}
\DeclareUnicodeCharacter{22A2}{\vdash}
\DeclareUnicodeCharacter{2090}{\ensuremath{{}_a}}
\DeclareUnicodeCharacter{A71B}{{}^\uparrow}
\DeclareUnicodeCharacter{A71C}{{}^\downarrow}
\DeclareUnicodeCharacter{27E6}{\llbracket}
\DeclareUnicodeCharacter{27E7}{\rrbracket}

\DeclareUnicodeCharacter{2460}{\textcircled{\small{1}}}
\DeclareUnicodeCharacter{2461}{\textcircled{\small{2}}}
\DeclareUnicodeCharacter{2462}{\textcircled{\small{3}}}
\DeclareUnicodeCharacter{2463}{\textcircled{\small{4}}}
\DeclareUnicodeCharacter{2464}{\textcircled{\small{5}}}
\DeclareUnicodeCharacter{2465}{\textcircled{\small{6}}}
\DeclareUnicodeCharacter{2466}{\textcircled{\small{7}}}
\DeclareUnicodeCharacter{24EA}{\textcircled{\small{0}}}

\newcommand{\unicoderightcircle}{\ensuremath{\RIGHTcircle}}
\DeclareUnicodeCharacter{25D1}{\unicoderightcircle}
\newcommand{\unicodeleftcircle}{\ensuremath{\LEFTcircle}}
\DeclareUnicodeCharacter{25D0}{\unicodeleftcircle}
\DeclareUnicodeCharacter{229B}{\circledast}

\newcommand{\unicodebbA}{\ensuremath{\mathbb{A}}}
\DeclareUnicodeCharacter{1D538}{\unicodebbA}
\newcommand{\unicodebbB}{\ensuremath{\mathbb{B}}}
\DeclareUnicodeCharacter{1D539}{\unicodebbB}
\newcommand{\unicodebbC}{\ensuremath{\mathbb{C}}}
\DeclareUnicodeCharacter{2102}{\unicodebbC}
\DeclareUnicodeCharacter{1D53B}{\ensuremath{\mathbb{D}}}
\DeclareUnicodeCharacter{2115}{\ensuremath{\mathbb{N}}}
\DeclareUnicodeCharacter{211D}{\ensuremath{\mathbb{R}}}
\DeclareUnicodeCharacter{1D543}{\ensuremath{\mathbb{L}}}
\newcommand\UnicodeBlackboardP{\ensuremath{\mathbf{P}}} \DeclareUnicodeCharacter{2119}{\UnicodeBlackboardP}
\DeclareUnicodeCharacter{211A}{\ensuremath{\mathbb{Q}}}
\DeclareUnicodeCharacter{1D544}{\ensuremath{\mathbb{M}}}
\DeclareUnicodeCharacter{1D54C}{\ensuremath{\mathbb{U}}}
\DeclareUnicodeCharacter{1D54D}{\ensuremath{\mathbf{V}}}
\DeclareUnicodeCharacter{1D54E}{\ensuremath{\mathbb{W}}}
\DeclareUnicodeCharacter{1D542}{\ensuremath{\mathbb{K}}}
\DeclareUnicodeCharacter{1D546}{\ensuremath{\mathbb{O}}}
\DeclareUnicodeCharacter{1D540}{\ensuremath{\mathbb{I}}}
\DeclareUnicodeCharacter{1D54A}{\ensuremath{\mathbb{S}}}
\DeclareUnicodeCharacter{1D53C}{\ensuremath{\mathbb{E}}}

\DeclareUnicodeCharacter{1D405}{\ensuremath{\mathbf{F}}}
\DeclareUnicodeCharacter{1D406}{\ensuremath{\mathbf{G}}}
\DeclareUnicodeCharacter{1D413}{\ensuremath{\mathbf{T}}}
\DeclareUnicodeCharacter{1D5E4}{\ensuremath{\mathbf{Q}}}

\newcommand{\unicodecalS}{\ensuremath{\mathcal{S}}}
\newcommand{\unicodecalT}{\ensuremath{\mathcal{T}}}
\newcommand{\unicodecalC}{\ensuremath{\mathcal{C}}}

\newcommand{\unicodecalX}{\ensuremath{\mathcal{X}}}
\newcommand{\unicodecalN}{\ensuremath{\mathcal{N}}}
\newcommand{\unicodecalE}{\ensuremath{\mathcal{E}}}
\DeclareUnicodeCharacter{1D4D4}{\unicodecalE}
\DeclareUnicodeCharacter{1D4D2}{\unicodecalC}

\DeclareUnicodeCharacter{1D4DE}{\mathcal{O}}
\DeclareUnicodeCharacter{1D4AA}{\mathcal{O}}
\DeclareUnicodeCharacter{210B}{\mathcal{H}}
\DeclareUnicodeCharacter{1D4E2}{\unicodecalS}
\DeclareUnicodeCharacter{1D4E3}{\unicodecalT}
\DeclareUnicodeCharacter{1D4E7}{\unicodecalX}
\DeclareUnicodeCharacter{1D4D0}{\ensuremath{\mathcal{A}}}
\DeclareUnicodeCharacter{1D4D1}{\ensuremath{\mathcal{B}}}
\DeclareUnicodeCharacter{1D4D6}{\ensuremath{\mathcal{G}}}
\DeclareUnicodeCharacter{1D4D7}{\ensuremath{\mathcal{H}}}
\DeclareUnicodeCharacter{1D4DB}{\ensuremath{\mathcal{L}}}

\DeclareUnicodeCharacter{1D4DD}{\unicodecalN}
\DeclareUnicodeCharacter{1D4B1}{\ensuremath{\mathcal{V}}}
\DeclareUnicodeCharacter{1D4E5}{\ensuremath{\mathcal{V}}}
\DeclareUnicodeCharacter{1D4E6}{\ensuremath{\mathcal{W}}}
\DeclareUnicodeCharacter{1D4E4}{\ensuremath{\mathcal{U}}}

\DeclareUnicodeCharacter{03B1}{\alpha}
\DeclareUnicodeCharacter{03B2}{\beta}
\DeclareUnicodeCharacter{03BC}{\mu}
\DeclareUnicodeCharacter{03B4}{\delta}
\DeclareUnicodeCharacter{03B5}{\varepsilon}
\DeclareUnicodeCharacter{03B7}{\eta}
\DeclareUnicodeCharacter{03BB}{\lambda}
\DeclareUnicodeCharacter{03C1}{\rho}
\DeclareUnicodeCharacter{03C8}{\psi}
\DeclareUnicodeCharacter{03C4}{\tau}
\DeclareUnicodeCharacter{03A8}{\Psi}
\DeclareUnicodeCharacter{03C3}{\sigma}
\DeclareUnicodeCharacter{03C6}{\varphi}
\DeclareUnicodeCharacter{03A6}{\Phi}
\DeclareUnicodeCharacter{03A3}{\Sigma}
\DeclareUnicodeCharacter{03D5}{\phi}
\DeclareUnicodeCharacter{03B8}{\theta}
\DeclareUnicodeCharacter{03C0}{\ensuremath{\pi}}
\DeclareUnicodeCharacter{0393}{\Gamma}
\DeclareUnicodeCharacter{0394}{\Delta}
\DeclareUnicodeCharacter{03BA}{\kappa}
\DeclareUnicodeCharacter{03BD}{\nu}
\DeclareUnicodeCharacter{03A9}{\Omega}
\DeclareUnicodeCharacter{03C9}{\omega}
\DeclareUnicodeCharacter{03D2}{\Upsilon}
\DeclareUnicodeCharacter{03C5}{\upsilon}
\DeclareUnicodeCharacter{22A0}{\boxtimes}
\DeclareUnicodeCharacter{1D412}{\ensuremath{\mathbf{S}}}

\DeclareFontFamily{U}{mathb}{\hyphenchar\font45}
\DeclareSymbolFont{mathb}{U}{mathb}{m}{n}
\DeclareMathSymbol{\sqbullet}{\mathbin}{mathb}{"0D}
\DeclareUnicodeCharacter{25A0}{\sqbullet}
\DeclareUnicodeCharacter{25AA}{\mathbin{\sqbullet}}

\DeclareUnicodeCharacter{2205}{\varnothing}
\DeclareUnicodeCharacter{2254}{\coloneq}
\DeclareUnicodeCharacter{25AB}{\blacksquare}
\DeclareUnicodeCharacter{03B3}{\gamma}

\DeclareUnicodeCharacter{112A9}{\mathbin{\pmb{;}}}

\newcommand{\hirayo}{\scaleobj{0.9}{\text{\usefont{U}{min}{m}{n}\symbol{'210}}}}
\DeclareUnicodeCharacter{3088}{\hirayo}
\DeclareFontFamily{U}{min}{}
\DeclareFontShape{U}{min}{m}{n}{<-> udmj30}{}

\newcommand\UnicodeWhiteRightPointingSmallTriangle{\triangleright}
\DeclareUnicodeCharacter{25B9}{\mathbin{\UnicodeWhiteRightPointingSmallTriangle}}
\newcommand\UnicodeWhiteDownPointingSmallTriangle{\triangledown}
\DeclareUnicodeCharacter{25BF}{\mathbin{\UnicodeWhiteDownPointingSmallTriangle}}
\newcommand\UnicodeWhiteUpPointingSmallTriangle{\scalemath{1}{-1}{{}^{\triangledown}}}
\DeclareUnicodeCharacter{25B5}{\mathbin{\UnicodeWhiteUpPointingSmallTriangle}}

\DeclareUnicodeCharacter{2080}{\ensuremath{{}_0}}
\DeclareUnicodeCharacter{2081}{\ensuremath{{}_1}}
\DeclareUnicodeCharacter{2082}{\ensuremath{{}_2}}
\DeclareUnicodeCharacter{2083}{\ensuremath{{}_3}}

\DeclareUnicodeCharacter{1D62}{\ensuremath{{}_i}}
\DeclareUnicodeCharacter{2C7C}{\ensuremath{{}_j}}
\DeclareUnicodeCharacter{02B3}{\ensuremath{{}^r}}
\DeclareUnicodeCharacter{02E1}{\ensuremath{{}^\ell}}
\DeclareUnicodeCharacter{1D48}{\ensuremath{{}^d}}
\DeclareUnicodeCharacter{1D50}{\ensuremath{{}^m}}
\DeclareUnicodeCharacter{1D58}{\ensuremath{{}^u}}
\DeclareUnicodeCharacter{209A}{\ensuremath{{}_p}}
\DeclareUnicodeCharacter{2096}{\ensuremath{{}_k}}

\DeclareUnicodeCharacter{2245}{\ensuremath{\cong}}
\DeclareUnicodeCharacter{2286}{\subseteq}

\DeclareUnicodeCharacter{22C5}{\cdot}
\DeclareUnicodeCharacter{25C3}{\ensuremath{\triangleleft}}
\DeclareUnicodeCharacter{25B9}{\ensuremath{\triangleright}}

\DeclareUnicodeCharacter{2217}{\ast}
\DeclareUnicodeCharacter{039E}{\Xi}
\DeclareUnicodeCharacter{2295}{\oplus}
\DeclareUnicodeCharacter{2297}{\otimes}
\DeclareUnicodeCharacter{214B}{\parr}
\DeclareUnicodeCharacter{2298}{\oslash}
\DeclareUnicodeCharacter{25C0}{\mathbin{\blacktriangleleft}}
\DeclareUnicodeCharacter{25C1}{\mathbin{\vartriangleleft}}
\DeclareUnicodeCharacter{22B3}{\mathbin{\triangleright}}
\DeclareUnicodeCharacter{22B2}{\mathbin{\triangleleft}}
\DeclareUnicodeCharacter{FF5C}{\mid}
\DeclareUnicodeCharacter{227A}{\mathbin{\prec}}
\DeclareUnicodeCharacter{227B}{\mathbin{\succ}}
\DeclareUnicodeCharacter{22A3}{\mathbin{\dashv}}
\DeclareUnicodeCharacter{219D}{\ensuremath{\leadsto}}
\DeclareUnicodeCharacter{2191}{\ensuremath{\uparrow}}
\DeclareUnicodeCharacter{1361}{\colon}

\DeclareUnicodeCharacter{29D1}{\mathrel{\multimapdotbothB}}
\DeclareUnicodeCharacter{29D2}{\mathrel{\multimapdotbothA}}
\DeclareUnicodeCharacter{22C4}{\mathbin{\diamond}}
\DeclareUnicodeCharacter{226B}{\mathrel{\gg}}
\DeclareUnicodeCharacter{25A1}{\mathbin{\square}}
\DeclareUnicodeCharacter{266F}{\sharp}

\DeclareUnicodeCharacter{2113}{\ell}
\DeclareUnicodeCharacter{2261}{\equiv}
\DeclareUnicodeCharacter{2099}{_n}
\DeclareUnicodeCharacter{2098}{_m}
\DeclareUnicodeCharacter{1D5AD}{\ensuremath{\mathsf{N}}}
\DeclareUnicodeCharacter{1D5E1}{\ensuremath{\mathbf{N}}}

\DeclareUnicodeCharacter{1D4DF}{\mathcal{P}}
\DeclareUnicodeCharacter{1D415}{\mathbf{V}}
\DeclareUnicodeCharacter{1D400}{\mathbf{A}}
\DeclareUnicodeCharacter{1D6C2}{\pmb{\alpha}}
\DeclareUnicodeCharacter{1D6C3}{\pmb{\beta}}
\DeclareUnicodeCharacter{1D6C4}{\pmb{\gamma}}
\DeclareUnicodeCharacter{1D6C5}{\pmb{\delta}}
\DeclareUnicodeCharacter{1D6C6}{\pmb{\epsilon}}
\DeclareUnicodeCharacter{1D6C8}{\pmb{\eta}}
\DeclareUnicodeCharacter{1D6DA}{\pmb{\omega}}

\newcommand\mydots{\makebox[0.6em][c]{.\hfil.\hfil.}}
\DeclareUnicodeCharacter{2026}{\mydots}
\DeclareUnicodeCharacter{226B}{\gg}

\DeclareUnicodeCharacter{22C9}{\ltimes}
\DeclareUnicodeCharacter{22CA}{\rtimes}

\usepackage[only,fatsemi]{stmaryrd}
\newcommand{\unicodeRelationalComposition}{\fatsemi}
\DeclareUnicodeCharacter{2A3E}{\unicodeRelationalComposition}
\DeclareUnicodeCharacter{2983}{\llparenthesis}
\DeclareUnicodeCharacter{2984}{\rrparenthesis}

\DeclareUnicodeCharacter{2014}{\,---\,} %
\usepackage{xcolor}

\definecolor{nordred}{HTML}{bf616a}
\definecolor{bordeaux}{HTML}{821529}
\definecolor{bluelink}{HTML}{003399}
\definecolor{nordred}{HTML}{bf616a}
\definecolor{nordblue}{HTML}{81a1c1}
\definecolor{norddarkblue}{HTML}{5e81ac}
\definecolor{nordgreen}{HTML}{a3be8c}
\definecolor{nordnight}{HTML}{4c566a}

\AtEndPreamble{
  \RequirePackage{hyperref}
  \RequirePackage{cleveref}
  \hypersetup{
    breaklinks = true,
    linktocpage,
    colorlinks = true,
    linkcolor = nordnight,
    citecolor = nordgreen,
    urlcolor = nordblue
  }
} %
\makeatletter
\newcommand{\nicelinktarget}[1]{\Hy@raisedlink{\hypertarget{#1}{}}}
\makeatother

\renewcommand\Set{\hyperlink{linkSet}{\mathbf{Set}}}

\usepackage[xcolor,no patch,hyperref,quotation,electronic]{knowledge}
\knowledgeconfigure{notion}
\knowledgestyle{notion}{color=nordnight}
\knowledgestyle{intro notion}{emphasize,color=nordnight}

\knowledge{notion}
| string diagrams for commutative magmoids
| String diagrams for commutative magmoids

\knowledge{notion}
| uniformity
| Uniformity
| uniformity axiom
| Uniformity axiom
| uniformly traced distributive multicategory
| uniformly traced distributive multicategories
| Uniformly traced distributive multicategory
| Uniformly traced distributive multicategories

\knowledge{notion}
| associating morphism
| Associating morphism
| associating morphisms
| Associating morphisms
| associating
| Associating

\knowledge{notion}
| distributive sesquilaw
| distributive sesquilaws
| Distributive sesquilaw
| Distributive sesquilaws
| sesquilaw
| sesquilaws
| Sesquilaw
| Sesquilaws

\knowledge{notion}
| associative
| Associative

\knowledge{notion}
| sesquilaw homomorphism
| sesquilaw homomorphisms
| Sesquilaw homomorphism
| Sesquilaw homomorphisms

\knowledge{notion}
| monoidal sesquilaw
| monoidal sesquilaws
| Monoidal sesquilaw
| Monoidal sesquilaws

\knowledge{notion}
| post-selection
| Post-selection

\knowledge{notion}
| normalized probabilistic channels
| normalized probabilistic channel
| Normalized probabilistic channels
| Normalized probabilistic channel
| Normalized kernels
| Normalized kernel
| normalized kernel
| normalized kernels

\knowledge{notion}
| Continuous normalized kernels
| Continuous normalized kernel
| continuous normalized kernel
| continuous normalized kernels

\knowledge{notion}
| bracketed division
| Bracketed division

\knowledge{notion}
| monoidal natural transformation
| monoidal natural transformations
| Monoidal natural transformation
| Monoidal natural transformations
| monoidal transformation
| monoidal transformations
| Monoidal transformation
| Monoidal transformations

\knowledge{notion}
| monad
| monads
| Monad
| Monads

\knowledge{notion}
| distribution
| Distribution
| distributions
| Distributions

\knowledge{notion}
| non-empty powerset monad
| Non-empty powerset monad
| non-empty powerset
| Non-empty powerset
| Affine powerset monad
| affine powerset monad
| Affine powerset
| affine powerset
\DeclareUnicodeCharacter{1D411}{\kl[non-empty powerset monad]{\mathbf{R}}} %

\knowledge{notion}
| non-empty relation
| non-empty relations
| Non-empty relation
| Non-empty relations
| affine relation
| affine relations
| Affine relation
| Affine relations

\knowledge{notion}
| subaffine relation
| subaffine relations
| Subaffine relation
| Subaffine relations

\knowledge{notion}
| partial affine relation
| partial affine relations
| Partial affine relation
| Partial affine relations

\knowledge{notion}
| monoidal monad
| monoidal monads
| Monoidal monad
| Monoidal monads

\knowledge{notion}
| monoidal non-associative monad
| monoidal non-associative monads
| Monoidal non-associative monad
| Monoidal non-associative monads

\knowledge{notion}
| natural transformation
| natural transformations
| Natural transformation
| Natural transformations

\knowledge{notion}
| symmetric monoidal categories
| symmetric monoidal category
| Symmetric monoidal categories
| Symmetric monoidal category

\knowledge{notion}
| distributive law
| distributive laws
| Distributive law
| Distributive laws

\knowledge{notion}
| monoidal distributive law
| monoidal distributive laws
| Monoidal distributive law
| Monoidal distributive laws

\knowledge{notion}
| almost distributive law
| almost distributive laws
| Almost distributive law
| Almost distributive laws
| almost-distributive law
| almost-distributive laws
| Almost-distributive law
| Almost-distributive laws

\knowledge{notion}
| term
| terms
| Term
| Terms

\knowledge{notion}
| interchange
| Interchange
| interchange axiom
| Interchange axiom

\knowledge{notion}
| affine magmoid
| Affine magmoid
| affine magmoids
| Affine magmoids

\knowledge{notion}
| affine
| Affine
| affine monad
| Affine monad

\knowledge{notion}
| relevant
| Relevant
| relevant monad
| Relevant monad

\knowledge{notion}
| copy-discard sesquilaw
| Copy-discard sesquilaw
| copy-discard sesquilaws
| Copy-discard sesquilaws

\knowledge{notion}
| sesquilaw magmoid
| Sesquilaw magmoid
| sesquilaw magmoids
| Sesquilaw magmoids

\knowledge{notion}
| correctness triple
| Correctness triple
| correctness triples
| Correctness triples

\knowledge{notion}
| posetal imperative category
| Posetal imperative category
| posetal imperative categories
| Posetal imperative categories

\knowledge{notion}
| posetal imperative multicategory
| Posetal imperative multicategory
| posetal imperative multicategories
| Posetal imperative multicategories

\knowledge{notion}
| full support
| Full support

\knowledge{notion}
| hypergraph
| Hypergraph
| hypergraph magmoid
| hypergraph magmoids
| Hypergraph magmoid
| Hypergraph magmoids

\knowledge{notion}
| guard combinators
| Guard combinators
| guard combinator
| Guard combinator

\knowledge{notion}
| command combinators
| Command combinators
| command combinator
| Command combinator

\knowledge{notion}
| monoidal category
| monoidal categories
| Monoidal category
| Monoidal categories

\knowledge{notion}
| sesquifunctor
| Sesquifunctor
| sesquifunctors
| Sesquifunctors

\knowledge{notion}
| premonoidal category
| Premonoidal category
| premonoidal categories
| Premonoidal categories
| premonoidality
| Premonoidality
| premonoidal
| Premonoidal

\knowledge{notion}
| central
| Central
| central morphism
| Central morphism
| central morphisms
| Central morphisms

\knowledge{notion}
| Symmetric premonoidal category
| Symmetric premonoidal categories
| symmetric premonoidal category
| symmetric premonoidal categories

\knowledge{notion}
| deterministic
| Deterministic
| deterministic morphism
| Deterministic morphism
| deterministic morphisms
| Deterministic morphisms

\knowledge{notion}
| total
| total morphism
| total morphisms
| Total
| Total morphism
| Total morphisms

\knowledge{notion}
| cocartesian multicategory
| Cocartesian multicategory
| cocartesian multicategories
| Cocartesian multicategories

\knowledge{notion}
| predistributive multicategory
| Predistributive multicategory
| predistributive multicategories
| Predistributive multicategories

\knowledge{notion}
| posetal distributive copy-discard multicategory
| Posetal distributive copy-discard multicategory
| posetal distributive copy-discard multicategories
| Posetal distributive copy-discard multicategories

\knowledge{notion}
| index
| indices
| Index
| Indices

\knowledge{notion}
| state
| states
| State
| States

\knowledge{notion}
| state combinator
| state combinators
| State combinator
| State combinators

\knowledge{notion}
| multicategory
| Multicategory
| Multicategories
| multicategories

\knowledge{notion}
| distributive multicategory
| distributive multicategories
| Distributive multicategory
| Distributive multicategories

\knowledge{notion}
| substitution
| Substitution
| substitutions
| Substitutions
| substituting
| Substituting

\knowledge{notion}
| guard
| guards
| Guard
| Guards

\knowledge{notion}
| predicate
| predicates
| Predicates
| Predicate

\knowledge{notion}
| predicate combinators
| Predicate combinators

\knowledge{notion}
| command
| commands
| Commands
| Command

\knowledge{notion}
| fresh
| Fresh

\knowledge{notion}
| alpha-equivalent
| alpha-equivalence
| Alpha-equivalent
| Alpha-equivalence

\knowledge{notion}
| variable
| variables
| Variable
| Variables

\knowledge{notion}
| traced distributive copy-discard multicategory
| Traced distributive copy-discard multicategory
| traced distributive copy-discard multicategories
| Traced distributive copy-discard multicategories

\knowledge{notion}
| traced distributive multicategory
| Traced distributive multicategory
| traced distributive multicategories
| Traced distributive multicategories

\knowledge{notion}
| traced predistributive copy-discard multicategory
| Traced predistributive copy-discard multicategory
| traced predistributive copy-discard multicategories
| Traced predistributive copy-discard multicategories

\knowledge{notion}
| predistributive copy-discard multicategory
| Predistributive copy-discard multicategory
| predistributive copy-discard multicategories
| Predistributive copy-discard multicategories

\knowledge{notion}
| distributive copy-discard multicategory
| Distributive copy-discard multicategory
| distributive copy-discard multicategories
| Distributive copy-discard multicategories

\knowledge{notion}
| variable substitution
| variable substitutions
| Variable substitution
| Variable substitutions

\knowledge{notion}
| posetal distributive signature
| posetal distributive signatures
| Posetal distributive signature
| Posetal distributive signatures

\knowledge{notion}
| posetal uniformity
| Posetal uniformity

\knowledge{notion}
| posetal distributive copy-discard category
| posetal distributive copy-discard categories
| Posetal distributive copy-discard category
| Posetal distributive copy-discard categories

\knowledge{notion}
| posetal uniform trace
| posetal uniform traces
| Posetal uniform traces
| Posetal uniform trace
| posetal uniform traced monoidal category
| Posetal uniform traced monoidal category
| posetal uniform traced monoidal categories
| Posetal uniform traced monoidal categories

\knowledge{notion}
| label substitution
| label substitutions
| Label substitution
| Label substitutions

\knowledge{notion}
| commutative Lawvere theory
| Commutative Lawvere theory
| commutative Lawvere theories
| Commutative Lawvere theories

\knowledge{notion}
| Lawvere theory
| Lawvere theories
| clone
| Clone
| clones
| Clones

\knowledge{notion}
| anchor
| anchors
| Anchor
| Anchors
| label
| Label
| Labels
| labels

\knowledge{notion}
| conditional
| conditionals
| Conditional
| Conditionals

\knowledge{notion}
| conditional composition
| conditional compositions
| Conditional composition
| Conditional compositions

\knowledge{notion}
| context
| Context
| contexts
| Contexts

\knowledge{notion}
| copy-discard category
| Copy-discard category
| copy-discard categories
| Copy-discard categories
| copy-discard
| Copy-discard

\knowledge{notion}
| copy-discard monoidal category
| Copy-discard monoidal category
| copy-discard monoidal categories
| Copy-discard monoidal categories

\knowledge{notion}
| copy-discard premonoidal category
| Copy-discard premonoidal category
| copy-discard premonoidal categories
| Copy-discard premonoidal categories
| premonoidal copy-discard category
| Premonoidal copy-discard category
| premonoidal copy-discard categories
| Premonoidal copy-discard categories

\knowledge{notion}
| commutative imperative category
| Commutative imperative category
| commutative imperative categories
| Commutative imperative categories

\knowledge{notion}
| imperative category
| Imperative category
| imperative categories
| Imperative categories

\knowledge{notion}
| imperative multicategory
| Imperative multicategory
| imperative multicategories
| Imperative multicategories

\knowledge{notion}
| multiquiver
| multiquivers
| Multiquiver
| Multiquivers

\knowledge{notion}
| tricocycloid
| tricocycloids
| Tricocycloid
| Tricocycloids
| symmetric tricocycloid
| symmetric tricocycloids
| Symmetric tricocycloid
| Symmetric tricocycloids

\knowledge{notion}
| normalized distributions magmoid
| normalized distribution magmoid
| normalization magmoid
| Normalized distributions magmoid
| Normalized distribution magmoid
| Normalization magmoid
| monoidal magmoid of normalized finitary distributions
| normalized finitary distribution magmoid

\knowledge{notion}
| finite normalized kernels
| finite normalized kernel
| Finite normalized kernels
| Finite normalized kernel

\knowledge{notion}
| left-relevant magmoid
| left-relevant magmoids
| Left-relevant magmoid
| Left-relevant magmoids

\knowledge{notion}
| quasitotal magmoid
| quasitotal magmoids
| Quasitotal magmoid
|  magmoids

\knowledge{notion}
| monoidal magmoid
| monoidal magmoids
| Monoidal magmoid
| Monoidal magmoids

\knowledge{notion}
| symmetric monoidal magmoid
| symmetric monoidal magmoids
| Symmetric monoidal magmoid
| Symmetric monoidal magmoids

\knowledge{notion}
| symmetric monoidal non-associative category
| Symmetric monoidal non-associative category
| symmetric monoidal non-associative categories
| Symmetric monoidal non-associative categories

\knowledge{notion}
| tensor schema
| tensor schemas
| Tensor schema
| Tensor schemas

\knowledge{notion}
| string diagram
| string diagrams
| String diagram
| String diagrams

\knowledge{notion}
| commuting magmoid
| commuting magmoids
| Commuting magmoid
| Commuting magmoids
| commutative magmoid
| commutative magmoids
| Commutative magmoid
| Commutative magmoids
| left commutative magmoid
| left commutative magmoids
| Left commutative magmoid
| Left commutative magmoids
\knowledge{notion}
| right commutative magmoid
| right commutative magmoids
| Right commutative magmoid
| Right commutative magmoids

\knowledge{notion}
| commutative non-associative monad
| commutative non-associative monads
| Commutative non-associative monad
| Commutative non-associative monads
| commuting non-associative monad
| commuting non-associative monads
| Commuting non-associative monad
| Commuting non-associative monads

\knowledge{notion}
| strict monoidal magmoid
| strict monoidal magmoids
| Strict monoidal magmoid
| Strict monoidal magmoids

\knowledge{notion}
| bimonoidally strict
| Bimonoidally strict
| strict
| Strict

\knowledge{notion}
| distributive signature
| distributive signatures
| Distributive signature
| Distributive signatures

\knowledge{notion}
| distributive category
| distributive categories
| Distributive category
| Distributive categories
| distributive copy-discard category
| distributive copy-discard categories
| Distributive copy-discard category
| Distributive copy-discard categories

\knowledge{notion}
| basic type
| basic types
| Basic type
| Basic types

\knowledge{notion}
| generator
| generators
| Generator
| Generators

\knowledge{notion}
| effectful triple
| effectful triples
| Effectful triple
| Effectful triples

\knowledge{notion}
| double signature
| double signatures
| Double signature
| Double signatures

\knowledge{notion}
| forgetful double signature
| forgetful double signatures
| Forgetful double signature
| Forgetful double signatures

\knowledge{notion}
| double signature morphism
| double signature morphisms
| Double signature morphism
| Double signature morphisms

\knowledge{notion}
| pinwheel double category
| pinwheel double categories
| Pinwheel double category
| Pinwheel double categories
| double category with pinwheels
| double categories with pinwheels
| Double category with pinwheels
| Double categories with pinwheels

\knowledge{notion}
| bigraph
| Bigraph
| bigraphs
| Bigraphs
| 2-graph
| 2-graphs
| 2-Graph
| 2-Graphs

\knowledge{notion}
| 2-graph morphism
| 2-graph morphisms

\knowledge{notion}
| 2-graph category
| 2-graph categories

\knowledge{notion}
| double category
| double categories
| Double category
| Double categories

\knowledge{notion}
| weighted polygraph
| weighted polygraphs
| Weighted polygraph
| Weighted polygraphs
| timed polygraph
| timed polygraphs
| Timed polygraph
| Timed polygraphs

\knowledge{notion}
| morphism of timed polygraphs
| Morphism of timed polygraphs
| timed polygraph morphism
| Timed polygraph morphism

\knowledge{notion}
| tilted bicategory signature
| tilted bicategory signatures
| Tilted bicategory signature
| Tilted bicategory signatures
| tilted bigraph
| Tilted bigraph
| tilted bigraphs
| Tilted bigraphs
| tilted 2-graph
| Tilted 2-graph
| tilted 2-graphs
| Tilted 2-graphs

\knowledge{notion}
| path
| paths
| Path
| Paths
| composable path
| composable paths

\knowledge{notion}
| signature
| signatures
| Signature
| Signatures

\knowledge{notion}
| congruence
| congruences
| Congruence
| Congruences

\knowledge{notion}
| symmetry
| symmetries
| Symmetry
| Symmetries

\knowledge{notion}
| profunctor
| profunctors
| Profunctor
| Profunctors

\knowledge{notion}
| left inclusion
| Left inclusion

\knowledge{notion}
| right inclusion
| Right inclusion

\knowledge{notion}
| left whiskering
| Left whiskering

\knowledge{notion}
| right whiskering
| Right whiskering

\knowledge{notion}
| inclusion and whiskering
| Inclusion and whiskering
| whiskering interacts with inclusion
| Whiskering interacts with inclusion

\knowledge{notion}
| whiskering
| Whiskering

\knowledge{notion}
| rewiring preserves whiskering
| Rewiring preserves whiskering
| whiskering preserves rewiring
| Whiskering preserves rewiring

\knowledge{notion}
| rewiring preserves interchange
| Rewiring preserves interchange

\knowledge{notion}
| rewiring preserves composition
| Rewiring preserves composition

\knowledge{notion}
| rewiring
| Rewiring 
| definition of rewiring

\knowledge{rewiring is an action}[]{notion}

\knowledge{notion}
| composition
| Composition
| term composition
| Term composition

\knowledge{notion}
| term composition is associative
| Term composition is associative

\knowledge{notion}
| term composition is unital
| Term composition is unital

\knowledge{notion}
| tensoring
| Tensoring

\knowledge{notion}
| arrow notation preterm
| Arrow notation preterm
| arrow notation preterms
| Arrow notation preterms

\knowledge{notion}
| arrow notation term
| Arrow notation term
| arrow notation terms
| Arrow notation terms

\knowledge{notion}
| almost monad
| almost monads
| Almost monad
| Almost monads

\knowledge{notion}
| category
| categories
| Category
| Categories

\knowledge{notion}
| polyquiver
| polyquivers
| Polyquiver
| Polyquivers

\knowledge{notion}
| monoidal almost distributive law
| Monoidal almost distributive law
| monoidal almost distributive laws
| Monoidal almost distributive laws

\knowledge{notion}
| symmetric monoidal magmoid
| Symmetric monoidal magmoid
| symmetric monoidal magmoids
| Symmetric monoidal magmoids

\knowledge{notion}
| observe-arrow notation
| Observe-arrow notation
| observe-arrow notation term
| Observe-arrow notation term
| observe-arrow notation terms
| Observe-arrow notation terms

\knowledge{notion}
| subdistribution
| Subdistribution
| subdistributions
| Subdistributions
| subdistributional
| substochastic kernel
| substochastic kernels
| Substochastic kernel
| Substochastic kernels

\knowledge{notion}
| rescaling
| Rescaling

\knowledge{notion}
| normalisation
| normalisations
| Normalisation
| Normalisations
| normalization
| normalizations
| Normalization
| Normalizations
| normalised
| Normalised

\knowledge{notion}
| continuous normalization
| Continuous normalization

\knowledge{notion}
| observe-arrow notation copy-discard-compare category of terms

\knowledge{notion}
| category of arrow notation terms

\knowledge{notion}
| copy-discard-compare categories
| copy-discard-compare category
| Copy-discard-compare categories
| Copy-discard-compare category

\knowledge{notion}
| Kleisli extension
| Kleisli extensions

\knowledge{notion}
| Kleisli category
| Kleisli categories

\knowledge{notion}
| partially additive
| partially additive monad
| Partially additive
| Partially additive monad

\knowledge{notion}
| sets
| category of sets
\renewcommand{\Set}{\kl[sets]{\mathsf{Set}}}

\knowledge{notion}
| powerset
| powerset monad

\knowledge{notion}
| category of relations
| Category of relations

\knowledge{notion}
| maybe
| maybe monad
| Maybe
| Maybe monad
\DeclareUnicodeCharacter{1D40C}{\kl[Maybe monad]{\mathbf{M}}}

\knowledge{notion}
| continuous maybe
| continuous maybe monad
| Continuous maybe
| Continuous maybe monad
\DeclareUnicodeCharacter{1D4DC}{\kl[continuous maybe monad]{\mathcal{M}}}

\knowledge{notion}
| partial stochastic kernels
| Partial stochastic kernels
| partial stochastic kernel
| Partial stochastic kernel

\knowledge{notion}
| category of partial functions
| Category of partial functions

\knowledge{notion}
| Markov category
| Markov categories

\knowledge{notion}
| Markov magmoid
| Markov magmoids

\knowledge{notion}
| discrete Markov magmoid
| discrete Markov magmoids
| Discrete Markov magmoid
| Discrete Markov magmoids

\knowledge{notion}
| exact observation
| Exact observation
| exact observations
| Exact observations

\knowledge{notion}
| discrete copy-discard magmoid
| discrete copy-discard magmoids
| Discrete copy-discard magmoid
| Discrete copy-discard magmoids
| Frobenius equation

\knowledge{notion}
| quasitotal
| Quasitotal
| quasitotality
| Quasitotality
| quasitotal morphism
| Quasitotal morphism
| quasitotal morphisms
| Quasitotal morphisms

\knowledge{notion}
| partial Markov category
| partial Markov categories
| Partial Markov category
| Partial Markov categories

\knowledge{notion}
| quasi-Markov category
| quasi-Markov categories
| Quasi-Markov category
| Quasi-Markov categories

\knowledge{notion}
| monoidal almost-distributive law
| monoidal almost-distributive laws
| Monoidal almost-distributive law
| Monoidal almost-distributive laws

\knowledge{notion}
| almost-distributive law
| almost-distributive laws
| Almost-distributive law
| Almost-distributive laws

\knowledge{notion}
| magmoid
| Magmoid
| magmoids
| Magmoids
| Unital magmoid
| unital magmoid
| Unital magmoids
| unital magmoids
\knowledge{notion}
| non-associative category
| Non-associative category
| non-associative categories
\knowledge{notion}
| monoidal non-associative category
| Monoidal non-associative category
| monoidal non-associative categories
| Monoidal non-associative categories

\knowledge{notion}
| Giry monad
\newcommand{\unicodecalD}{\ensuremath{\kl[Giry monad]{\mathcal{D}}}}
\DeclareUnicodeCharacter{1D4D3}{\unicodecalD}

\knowledge{notion}
| finitary distribution monad
| distribution monad
| Finitary distribution monad
| Distribution monad
\DeclareUnicodeCharacter{1D403}{\kl[distribution monad]{\mathbf{D}}} %

\knowledge{notion}
| discrete subdistribution
| discrete subdistributions
| discrete subdistribution monad

\knowledge{notion}
| category of discrete stochastic channels
| Category of discrete stochastic channels

\knowledge{notion}
| support
| Support

\knowledge{notion}
| Dirac distribution

\knowledge{notion}
| measurable subdistribution
| measurable subdistributions
| measurable subdistribution monad

\knowledge{notion}
| standard Borel spaces
| Standard Borel spaces

\knowledge{notion}
| quantale-valued sets

\knowledge{notion}
| assertion-correctness triple
| assertion-correctness triples
| Assertion-correctness triple
| Assertion-correctness triples

\knowledge{notion}
| state-correctness triple
| state-correctness triples
| State-correctness triple
| State-correctness triples

\knowledge{notion}
| predicate-correctness triple
| predicate-correctness triples
| Predicate-correctness triple
| Predicate-correctness triples

\knowledge{notion}
| assertion-incorrectness triple
| assertion-incorrectness triples
| Assertion-incorrectness triple
| Assertion-incorrectness triples

\knowledge{notion}
| state-incorrectness triple
| state-incorrectness triples
| State-incorrectness triple
| State-incorrectness triples

\knowledge{notion}
| predicate-incorrectness triple
| predicate-incorrectness triples
| Predicate-incorrectness triple
| Predicate-incorrectness triples

\knowledge{notion}
| relational assertion-correctness triple
| relational assertion-correctness triples
| Relational assertion-correctness triple
| Relational assertion-correctness triples

\knowledge{notion}
| relational state-correctness triple
| relational state-correctness triples
| Relational state-correctness triple
| Relational state-correctness triples

\knowledge{notion}
| relational predicate-correctness triple
| relational predicate-correctness triples
| Relational predicate-correctness triple
| Relational predicate-correctness triples

\knowledge{notion}
| relational assertion-incorrectness triple
| relational assertion-incorrectness triples
| Relational assertion-incorrectness triple
| Relational assertion-incorrectness triples

\knowledge{notion}
| relational state-incorrectness triple
| relational state-incorrectness triples
| Relational state-incorrectness triple
| Relational state-incorrectness triples

\knowledge{notion}
| relational predicate-incorrectness triple
| relational predicate-incorrectness triples
| Relational predicate-incorrectness triple
| Relational predicate-incorrectness triples

\knowledge{notion}
| extension
| extensions
| Extension
| Extensions

\knowledge{notion}
| non-associative monad
| non-associative monads
| Non-associative monad
| Non-associative monads

\knowledge{notion}
| copy-discard magmoid
| copy-discard magmoids
| copy discard magmoid
| copy discard magmoids
| Copy-discard magmoid
| Copy-discard magmoids
| Copy discard magmoid
| Copy discard magmoids

\IfFileExists{noappendix.token}%
{\usepackage[appendix=strip,bibliography=common]{apxproof}}%
{\usepackage[bibliography=common]{apxproof}}

\theoremstyle{plain}
\newtheorem{theorem}{Theorem}[subsection]

\theoremstyle{definition}
\newtheorem{definition}[theorem]{Definition}

\newtheorem{algorithm}[theorem]{Algorithm}

\theoremstyle{remark}
\newtheorem{remark}[theorem]{Remark}

\definecolor{nord0}{RGB}{46, 52, 64}
\definecolor{nord1}{RGB}{59, 66, 82}
\definecolor{nord2}{RGB}{67, 76, 94}
\definecolor{nord3}{RGB}{76, 86, 106}
\definecolor{nord4}{RGB}{216, 222, 233}
\definecolor{nord5}{RGB}{229, 233, 240}
\definecolor{nord6}{RGB}{236, 239, 244}
\definecolor{nord7}{RGB}{143, 188, 187}
\definecolor{nord8}{RGB}{136, 192, 208}
\definecolor{nord9}{RGB}{129, 161, 193}
\definecolor{nord10}{RGB}{94, 129, 172}
\definecolor{nord11}{RGB}{191, 97, 106}
\definecolor{nord12}{RGB}{208, 135, 112}
\definecolor{nord13}{RGB}{235, 203, 139}
\definecolor{nord14}{RGB}{163, 190, 140}
\definecolor{nord15}{RGB}{180, 142, 173}

\usepackage{listings}

\lstdefinelanguage{Racket}{
  sensitive = true,
  alsoletter = {<,>,-,'},
  keywords={if, equal?, rDo, lDo, accDo, leftDo, rightDo, Norm, frontDoor, backDoor, <-},
  otherkeywords={==},
  keywords = [2]{match, @get, @post, @put, @delete, require, define, do,  Intervene, WithModel, Setting, To, In, intervene, Identify, withModel, setting, to, in, return, observe, ensure, otherwise, call, lang, racket, define-syntax, syntax-rules},
  keywords = [3]{'left, 'middle, 'right, '(),  uniform, uniformDoor, host, observeLeft, prevalence, test, uncertainty, 'gene, 'nogene, 'smoker, 'nonsmoker, 'tar, 'notar, list, <-},
  keywordstyle={\bfseries\color{nord3}},%
  keywordstyle=[2]\color{nord12},%
  keywordstyle=[3]{\color{nord3}},%
  keywordstyle=[4]\color{nord15},
  numbers=none, 
  basicstyle={\small\ttfamily\color{nord0}},
  columns=flexible,
  keepspaces=true,
  numberstyle={\tiny\ttfamily},
  stepnumber=1,
  numbersep=6pt,
  showstringspaces=false,
  breaklines=true,
  frame=single,
  frameround=tttt,
  comment=[l]{>},
  morecomment=[l]{;},
  commentstyle={\color{nord15}\ttfamily\itshape},
  stringstyle={\color{nord14}\ttfamily},
  morestring=[b]",
  literate={†}{$\smash{{}^{\dagger}}$}1 {λ}{$\lambda$}1 {η}{$\eta$}1
}

\allowdisplaybreaks

\title[Causal Probabilistic Programming via Magmadic Do-Notation]%
{Causal Probabilistic Programming\\via Magmadic Do-Notation}
\author{Mario Rom\'an}
\date{\today}

\makeatletter
\def\@copyrightspace{\relax}
\makeatother

\begin{document}

\begin{abstract}
  We introduce a do-notation metalanguage for causal probabilistic programming. The metalanguage is based on magmads: non-associative monads. We derive causal probabilistic programming constructs from non-associativity and the primitives of probabilistic programming.
\end{abstract}
\maketitle

\section{Introduction}

\subsection{Example —
  Simpson's paradox in observational clinical data}
\label{sec:example-simpson}

Imagine we have two treatments, named $A$ and $B$, for the same disease, $X$. From the observational data, $A$ seems slightly less successful than $B$: they achieve remission in $81.3\%$ and $83.8\%$ of the cases, respectively.

\addtolength{\abovecaptionskip}{-0.5em}
\addtolength{\belowcaptionskip}{0.5em}
\begin{figure}[h]
  \begin{tabular}{l|r|r}
    & \emph{Treatment} A & \emph{Treatment}  B \\ \hline
\emph{Illness $X$} & $81.3\%$ $(61 : 14)$ & $83.8\%$ $(88 : 17)$ \\ \hline
\emph{Variant $X_1$} & $93.3\%$ $(28 : 02)$ & $86.6\%$ $(78 : 12)$ \\
\emph{Variant $X_2$} & $73.3\%$ $(33 : 12)$ & $66.6\%$ $(10 : 05)$ 
\end{tabular}
\caption{Remission odds, by treatment and variant.\label{fig:remission}}
\end{figure}

However, we observe that clinical practice is to subdivide the illness in two variants: $X_1$ and $X_2$. After segregating patients (\Cref{fig:remission}), we find that treatment $A$ is separately more successful for $X_1$—with a success rate of $93.3\%$ versus $86.6\%$—and for $X_2$—with a success rate of $73.3\%$ versus $66.6\%$. The explanation for this apparent paradox is that the choice of treatment was being influenced by the variant—while the $X_1$ variant was mostly treated with $B$, the more difficult and rare $X_2$ variant was being systematically sent for treatment $A$.

This is \emph{Simpson's paradox}, a failure to account for a confounding variable in the observational data. Randomised trials avoid this issue: e.g., if we had truly randomized the application of $A$ or $B$, we would have been able to compare both under equal conditions.\footnote{This is not an entirely fictional scenario. In 1986, an observational study by Charig et al.~\cite{charig1986comparison} reported the relative success of different modalities of kidney stone surgery. Open surgery seemed slightly less successful than a modality of closed surgery, but this was a real-world instance of \emph{Simpson's paradox}, as noted in 1994 by Julious and Mullee \cite{julious1994confounding}, who concluded that ``randomised trials are therefore necessary to demonstrate any treatment effect.'' Similar real-world examples exist for many causal inference problems \cite{pearl2009causality}.}

But, do we always \emph{need} randomised trials to estimate treatment effect? 
\clearpage

\subsection{Example —
  causal reasoning with observational data}

In particular, in this example, we can estimate a randomised control trial from the given data.
\begin{enumerate}
  \item We can estimate the prevalence of the $X_1$ and $X_2$ variants. We have $120$ patients with the $X_1$ variant and $60$ patients with the $X_2$ variant (\Cref{fig:remission}).
  \item We can weight the success of the treatment $A$ on each variant by the actual probability of each, $28/30 · 120/180 + 33/45 · 60/180 = 13/15$.
  \item Analogously, we can weight the success of treatment $B$ by the actual probabilities, $78/90 · 120/180 + 10/15 · 60/180 = 12/15$.
  \item As a consequence, a $1{:}1$ randomised control trial of $A$ and $B$ with $30$ patients would expect $13/15$ to achieve remission with $A$ and $12/15$ to achieve remission with $B$.
\end{enumerate}

Thus, we have a mathematically valid estimator for a randomized control trial; and no randomized controlled trial was necessary for this estimate. Although the problem of finding estimators under causal assumptions is decidable and solved \cite{shpitser08a,tian2010identifying}, it remains unintuitive to reason about causality \cite{pearl2009causality}.

This manuscript introduces a syntax and a denotational semantics for a simple form of \emph{causal probabilistic programming}: a programming paradigm for causal inference built over \emph{probabilistic programming}.

\subsection{Probabilistic programming}

Probabilistic programming is a programming paradigm where statistical models are first-class citizens of the language; this feature facilitates a syntax for \emph{probabilistic inference}: the problem of estimating a conditional probability from observational data and a statistical model. 

While mathematically solved, probabilistic inference is conceptually difficult to apply correctly: probabilistic programming addresses this problem by allowing the declarative specification of probabilistic problems---it uses an \lstinline[language=Racket]|observe| operator that imposes a conditional constraint on a distribution (see, for instance, \Cref{fig:probabilistic-programming}).

\begin{figure}[h] 
\begin{minipage}{0.75\textwidth}
\begin{lstlisting}[language=Racket,numbers=none] 
(do (variant treatment outcome) <- data
    () <- (observe treatment 'A)
    return (outcome))
>>> '(((success) 61/75) 
>>>   ((failure) 14/75))

(do (variant treatment outcome) <- data
    () <- (observe treatment 'B)
    return (outcome))
>>> '(((success) 88/105) 
>>>   ((failure) 17/105))
\end{lstlisting}
\end{minipage}
\caption{Naive conditional estimation of treatment outcome.\label{fig:probabilistic-programming}}
\end{figure}

However, a naive application of probabilistic inference would fall prey to the confounding variable problem in \Cref{sec:example-simpson} (see the results in \Cref{fig:probabilistic-programming}). The query in \Cref{fig:probabilistic-programming} asks \emph{``after observing that I have been assigned to treatment A (or B), what are the odds of remission?''} Note that this is different from asking \emph{``what are the odds of remission if I undergo treatment A (or B)?''} 

Naively conditioning on observations does not answer this second question, for what we need is a causal model that explains out the bias introduced by deciding treatment based on variant. Endowed with a causal model, causal inference is solvable~\cite{shpitser08a}, but still conceptually difficult. This conceptual difficulty is eased by the syntax of \emph{causal probabilistic programming}.

\subsection{Causal probabilistic programming}
\emph{Causal probabilistic programming} allows us to declaratively specify problems of causal inference. Causal inference is the problem of estimating effects from observational data and a causal model \cite{pearl2009causality}. 
As an example, consider that we want to estimate the effect of a 1:1 randomized control trial for the data in \Cref{sec:example-simpson}. \Cref{fig:validquery} shows the query code: it assigns a treatment arm uniformly and estimates the causal effect of the treatment.

\begin{figure}[h]
  \begin{minipage}{0.75\textwidth}
\begin{lstlisting}[language=Racket,numbers=none]
(do (arm) <- (uniform '(A) '(B))
    (outcome) <- (intervene data
                  withModel (do 
                    variant <- ()
                    treatment <- (variant)
                    outcome <- (variant treatment)
                    return (variant treatment outcome))
                  setting (treatment) to (arm) 
                  in (outcome))
    return (arm outcome))
>>> '(((open success) 634983/1525400)
>>>  ((open failure) 127717/1525400)
>>>  ((closed success) 6231/16000)
>>>  ((closed failure) 1769/16000))
\end{lstlisting}
\end{minipage}
\caption{Causal estimation of a randomized control trial.\label{fig:validquery}}
\end{figure}

Causal inference is achieved via the \lstinline[language=Racket]|intervene|{} operator: it takes a distribution and a causal model and yields the intervened distribution—or an exception when impossible. \lstinline[language=Racket]|Intervene| is not a primitive: it is internally rewritten into non-associative sequencing and \lstinline[language=Racket]|observe|. For instance, \Cref{fig:validquery} becomes \Cref{fig:probrewriting}.

\begin{figure}[h]
  \begin{minipage}{0.75\textwidth}
\begin{lstlisting}[language=Racket,numbers=none]
(do (arm) <- (uniform '(A) '(B))
    (outcome) <- (do 
      (variant) <- (do 
        (variant1 treatment1 outcome1) <- data
        return (variant1))
      (outcome) <- (do 
        (variant2 treatment2 outcome2) <- data
        () <- (observe arm treatment2)
        () <- (observe variant variant2)
        return (outcome2))
      return (outcome))
    return (arm outcome))
\end{lstlisting}
\end{minipage}
\caption{Probabilistic rewriting of the causal estimation.\label{fig:probrewriting}}
\end{figure}

From this point of view, the primitives of our syntax are the (non-associative) \lstinline[language=Racket]|do| and the \lstinline[language=Racket]|observe| primitive from normalized probabilistic programming. Next section discusses the non-associative \lstinline[language=Racket]|do| primitive \cite{dilavore2026normalized} (\Cref{sec:non-associative-do-notation}); last section recasts an identifiability algorithm~\cite{shpitser08a} in terms of \lstinline[language=Racket]|do| and \lstinline[language=Racket]|observe| (\Cref{sec:identifiability}).

\begin{remark}
  The causal model can be itself declared using a form of single-output anonymous do-notation, as we do in \Cref{fig:validquery}. It specifies a semimarkovian model.
\end{remark}

\section{Magmadic Do-notation}

\label{sec:non-associative-do-notation}

Do-notation is right-associative by default \cite{moggi1991notions,wadler92}. We instead employ left-associative do-notation. We take semantics in the \kl{magmad}—the non-associative monad—of \kl{normalized distributions}~(\Cref{def:magmad-of-normalized-distributions}, see \cite{dilavore2026normalized}).

\subsection{Magmads}
Magmads are the non-associative counterpart of monads.\footnote{More precisely, we use \emph{unital} magmads and call them \emph{magmads} thoughout this text.} We introduce here Set-based magmads in \emph{relative form} (or \emph{Kleisli form}): magmads in terms of \emph{unit} and \emph{bind}.

\begin{definition}[Magmad]
  \label[definition]{def:magmad-relative}
  \AP A \emph{(unital)} \intro{magmad}, $(𝐓,η^{𝐓},β^{𝐓})$, consists of an assignment on sets, $𝐓X$ for each set $X$, a natural family of functions $\smash{η^{𝐓}} ፡ X → 𝐓X$ (called \emph{unit}) and a natural family of functions $\smash{β^{𝐓}} ፡ 𝐓 X × (X → 𝐓 Y) → 𝐓 Y$ (called \emph{bind}), satisfying the following two axioms.
  \begin{enumerate}
    \item $β^{𝐓}(η^{𝐓}(x))(λ x. f(x)) = f(x)$, left unitality; and
    \item $β^{𝐓}(t)(λ x. η^{𝐓}(x)) = t$, right unitality.
  \end{enumerate}
\end{definition}

\begin{remark}
  A \kl{monad} is a \kl{magmad} additionally satisfying associativity,
  \begin{enumerate}
    \item[(3)] $β^{𝐓}(t)(λ x. β^{𝐓}(f(x))(g)) = β^{𝐓}(β^{𝐓}(t)(λ x . f(x)))(λ y . g(y))$.
  \end{enumerate}
\end{remark}

\subsection{Magmadic do notation}

Because we do not have a monad but only a magmad, the usual monadic programming language \cite{moggi1991notions} can take two meanings depending on where we associate by default. Moreover, nested \lstinline[language=Racket]|do| does not reduce to a single \lstinline[language=Racket]|do|, as it happens with its associative counterpart.

\newsavebox{\codeTwo}
\begin{lrbox}{\codeTwo}
\begin{minipage}{0.35\textwidth}
\begin{lstlisting}[language=Racket]
(do (x ...) <- m1
    (y ...) <- m2
    rest ...)
\end{lstlisting}
\end{minipage}
\end{lrbox}

\newsavebox{\codeTwoR}
\begin{lrbox}{\codeTwoR}
\begin{minipage}{0.45\textwidth}
    \begin{lstlisting}[language=Racket]
(do (x ... y ...) <- 
       (do (x ...) <- m1
           (y ...) <- m2
           return (x ... y ...))
    rest ...)
\end{lstlisting}
  \end{minipage}
\end{lrbox}

\newsavebox{\codeThree}
\begin{lrbox}{\codeThree}
\begin{minipage}{0.35\textwidth}
    \begin{lstlisting}[language=Racket]
(do (x ...) <- m
    return (y ...))
\end{lstlisting}
\end{minipage}
\end{lrbox}

\newsavebox{\codeFour}
\begin{lrbox}{\codeFour}
\begin{minipage}{0.35\textwidth}
    \begin{lstlisting}[language=Racket]
(do return (y ...))
\end{lstlisting}
\end{minipage}
\end{lrbox}

\newsavebox{\codeFive}
\begin{lrbox}{\codeFive}
\begin{minipage}{0.35\textwidth}
    \begin{lstlisting}[language=Racket]
(do (x ...) <- m1
    (y ...) <- m2
    return (z ...))
\end{lstlisting}
\end{minipage}
\end{lrbox}

\begin{definition}[Magmadic do notation]
  Magmadic do notation is defined recursively by the following rules. The three first rules \emph{(i,ii,iii)} deal with the three base cases. The last rule \emph{(iv)} groups two lines, using $m = β(m_1; λ x. β(m_2; λ y. η(x,y)))$.
  \begin{align*}
  \usebox{\codeFour}  \quad & \overset{(i)}{=} \quad η(y_1,...,y_m) \\
  \usebox{\codeThree} \quad & \overset{(ii)}{=} \quad \begin{aligned}
    & β(m; λ(x_1,...,x_n). \\
    & \quad η(y_1,...,y_m)). 
  \end{aligned}\\
  \usebox{\codeFive} \quad & \overset{(iii)}{=} \quad \begin{aligned}
    & β(m_1; λ(x_1,...,x_n). \\
    & \quad β(m_2; λ(y_1,...,y_m). \\
    & \quad \quad η(z_1,...,z_k)))
  \end{aligned} \\
  \usebox{\codeTwo}   \quad & \overset{(iv)}{=} \quad \usebox{\codeTwoR}
  \end{align*}
\end{definition}

\subsection{Magmad of normalized distributions}

We recall the \emph{magmad}, or \emph{non-associative monad}, of normalized distributions~\cite{dilavore2026normalized}. A normalized distribution is either an empty formal sum—meaning it ``fails'', or that it is undefined—or a full distribution. Normalized distributions provide normalized stochastic semantics, but they do not form a monad, only a magmad.

\begin{definition}[Magmad of normalized distributions]
  \label[definition]{def:magmad-of-normalized-distributions}
  A \emph{normalized distribution} over a set is a finite formal sum over its elements whose coefficients are positive and add up to exactly $1$ or $0$. The \emph{finitary normalized distribution magmad}, $𝗡 ፡ \Set → \Set$, assigns, to each set, the set of normalized distributions over it,
  \[
  𝗡(X) = \left\{
    \sum_{i=0}^{n} λ_i \ket{x_i} \;\middle|\;
    x_i ∈ X,\, λ_i ∈ \mathbb{R}^{+}\!,\; \sum_{i=0}^{n} λ_i = 1 \mbox{ or } 0
  \right\}.
  \]
  Its unit, $\eta^{𝗡} ፡ X → 𝗡X$, is defined by $η^{𝗡}(x) = 1\ket{x}$. Its binding, $β^{𝗡} ፡ 𝗡X × (X → 𝗡Y) → 𝗡Y$, is defined by 
  \[β^{𝗡}\left( \sum_{i=0}^{n} λ_i \ket{x_i} ; \left[ x_i \mapsto \sum_{j=0}^{m_i} ρ_{ij} \ket{y_j} \right] \right) = \sum_{i=0}^{n} \sum_{j=0}^{m_i} \frac{λ_i ρ_{ij}\ket{y_j}}{{\sum_{i=0}^{n} \sum_{j=0}^{m_i} λ_i ρ_{ij}}}.\]
  Note that, whenever ${\sum_{i=0}^{n} \sum_{j=0}^{m_i} λ_i ρ_{ij}} = 0$, then there must be no summands and the value of the binding is the empty formal sum, $0$.
\end{definition}

\section{Identifiability}
\label{sec:identifiability}
\nosectionappendix

\Cref{sec:conditionals} derives conditional distributions from the basic primitives; \Cref{sec:identifiability-algorithms} derives the two main components of the algorithm.

\subsection{Conditionals from observations}
\label{sec:conditionals}

The conditional distribution $P(x|y)$ is traditionally computed by Bayes rule: $$P(x|y) = \frac{P(x)P(y|x)}{\sum\nolimits_{x'} P(x')P(y|x')}.$$
Note that this is precisely the normalization of the distribution $P(x)P(y|x) = P(x,y)$ constrained to a particular value of $y$. In terms of our probabilistic language, this is the same as computing $P(x,y)$ and then observing a particular value for $y$.

More generally, the value of $P(a_1,...,a_n\ |\ b_1,...,b_m)$ is obtained by computing $P(a_1,...,a_n,b_1,...,b_m)$ and observing particular values for the $b_i$ variables. In the following formula, $v_i$ refers to a temporary variable for the i-th output of the distribution; $v_{bi}$ refers to the temporary variable corresponding to $b_i$.
\[
\begin{minipage}{0.4\textwidth}
\begin{lstlisting}[language=Racket]
  p(a1 ... an | b1 ... bm)
\end{lstlisting}
\end{minipage}
 \quad = \quad
\begin{minipage}{0.4\textwidth}
      \begin{lstlisting}[language=Racket]
(do (v1 ... vk) <- p
    () <- (observe b1 vb1)
    ...
    () <- (observe bm vbm)
    return (a1 ... an))
\end{lstlisting}
\end{minipage}\]

\subsection{Identifiability algorithms}
\label{sec:identifiability-algorithms}

The \textbf{Identify} algorithm checks if a set of variables $C$ can be identified inside of a single confounded component $T$ containing them, $C ⊆ T$. It is based on Shpitser and Pearl's identify algorithm \cite{shpitser08a}, but differs in that \emph{(i)} it does not use multiplications or divisions, but only do-notation; \emph{(ii)} it computes conditionals separately, using \lstinline[language=Racket]|observe|; and \emph{(iii)} it outputs the selected variable as a single separated output, $x ∈ C$, instead of outputting all variables at once as in the original algorithm—this is helpful for the ID algorithm.

\begin{definition}[Causal model]
  A \emph{(semimarkovian) causal model} consists of a directed acyclic graph with a subset of \emph{visible} variables, such that each non-visible variable does not have any parents and is the parent of exactly two visible variables.
\end{definition}

\begin{algorithm}[Identify algorithm, c.f. {{\cite{shpitser08a}}}]
  Let $G$ be a causal model inducing a topological partial order $(≤)$. Let $T$ be a confounded component; let $C ⊆ T$ be a subset of it; and let $x ∈ C$ be a variable. Let $q$ be a distribution over $T$.

  The following algorithm computes $\mathrm{Identify}(q, G, T, C, x)$.
  \begin{enumerate}
    \item Compute $A = \mathrm{Ancestors}(C)_{G(T)}$, the ancestors of $C$ in the subgraph given by the component $T$.
    \item If $A = C$, then return the conditional distribution $q(x \mid \{c < x \mid c ∈ C\})$.
    \item If $A ≠ C$ and $A = T$, then fail.
    \item If $C \subset A \subset T$, then let $T'$ be the confounded component of $C$ in $G(A)$.
    \begin{enumerate}
      \item Compute a conditional distribution (by \Cref{sec:conditionals}) over the subset using the fact that it is a separated confounded component.
      {\[\begin{minipage}{0.6\textwidth}
      \begin{lstlisting}[language=Racket]
 q' = (do t1' <- q(t1' | {a : a < t1'})
          ...
          tn' <- q(tn' | {a : a < tn'})
          return (t1' ... tn'))
\end{lstlisting}
        \end{minipage}\]}
      \item Recursively apply $\mathrm{Identify}(q', G(T'), T', C, x)$.
    \end{enumerate}
  \end{enumerate}
\end{algorithm}

\begin{algorithm}[ID algorithm]
  The following algorithm computes $\mathrm{ID}(p, G, T, S)$.
  \begin{enumerate}
    \item Compute the ancestors of $S$ in the graph where all the variables in $T$ are removed, $D = \mathrm{Ancestors}(S)_{G(V - T)}$.
    \item For each variable $d_i ∈ D$, let $S_i$ be its confounded component.
    \item For each variable $d_i ∈ D$, let $D_i$ be its confounded component in the graph $G(D)$, restricted to variables in $D$.
    \item For each variable $d_i ∈ D$, let $QS_i$ be computed as a conditional distribution (by \Cref{sec:conditionals}) on a separated confounded component.
    {\[\begin{minipage}{0.6\textwidth}
      \begin{lstlisting}[language=Racket]
  QSi = (do si1 <- p(si1 | {v | v < si1})
            ...
            sim <- p(sim | {v | v < sim})
            return (si1 ... sim))
\end{lstlisting}
        \end{minipage}\]}
    \item For each variable in $D$, identify it using a call to the main Identify algorithm, $\mathrm{Identify}(QS_i, G, S_i, D_i, d_i)$.
    {\[\begin{minipage}{0.6\textwidth}
      \begin{lstlisting}[language=Racket]
  q = (do d1 <- Identify(QS1, G, S1, D1, d1)
          ...
          dn <- Identify(QSn, G, Sn, Dn, dn)
          return (s1 ... sn))
\end{lstlisting}
        \end{minipage}\]}
  \end{enumerate}
\end{algorithm}

\begin{remark}
  Shpitser and Pearl have shown that this algorithm is both sound—i.e., when it succeeds, it computes a correct estimator of the intervened distribution—and complete—i.e., it only fails when it is impossible to provide a correct estimator \cite{shpitser08a}. However, they have done so assuming a particular semantics: that of probability distributions.
\end{remark}

\subsection{Identifiability as a derived construct}

We introduce a single construct to magmadic do notation. It represents the result of modeling a distribution $p$ with a causal model $G$ and, after performing an intervention that sets some variables $(v_1,...,v_n)$ to some values $(x_1,...,x_n)$, identifying the value of some output variables $(y_1,...,y_m)$. We use the previous identifiability algorithm.

\begin{figure}[ht]
  \[\begin{minipage}{0.38\textwidth}
\begin{lstlisting}[language=Racket]
(intervene p withModel G
 setting (v ...) to (x ...) 
 in (y ...))
\end{lstlisting}
\end{minipage}\ \ =\ \begin{aligned}
  \mathrm{ID}(p,G,& \{v_1,...,v_n\},\\ &\{y_1,...,y_m\})(x_1,...,x_n)
\end{aligned}\]
\caption{Definition of the ``intervene'' construct.}
\end{figure}

In this way, every instance of the ``intervene'' construct is systematically rewritten into the output of the ID algorithm, which contains a probabilistic program estimating the intervention.\footnote{A Racket implementation of the algorithms described here is publicly available \cite{roman2026do}.  Additionally, the Appendix contains two more examples of automatic rewriting using the identifiability algorithm.}

\begin{toappendix}

\section{Further examples}

\subsection{Front-door criterion}
The paradigmatic problem of causal inference estimates the effect of smoking on cancer from some (fake) data suggesting a protective impact. The estimation uses an intermediate observed variable (\emph{tar in the lungs}) and this technique is usually known as the \emph{front-door criterion} \cite{pearl2009causality}. The query in \Cref{fig:smoking-example} asks, assuming this (fake) data: what would the incidence of cancer be if $5\%$ of the population were to smoke?

\begin{figure}[ht]
  \begin{minipage}{0.8\textwidth}
    \begin{lstlisting}[language=Racket]
(define survey (distribution-table
     ['(smoker tar nocancer)     323]
     ['(smoker tar cancer)        57]
     ['(nonsmoker tar nocancer)    1]
     ['(nonsmoker tar cancer)     19]
     ['(smoker notar nocancer)    18]
     ['(smoker notar cancer)       2]
     ['(nonsmoker notar nocancer) 38]
     ['(nonsmoker notar cancer)  342]))

; Which would the incidence be with a 5%
(define (incidence-from-habits habits)
  (intervene survey withModel 
    (do gene <- ()
        smoking <- (gene)
        tar <- (smoking)
        cancer <- (gene tar)
        return (smoking tar cancer))
   setting (smoking) to (habits) in (cancer)))

(do (newHabits) <- (distribution
                    ['(smoker)     5/100]
                    ['(nonsmoker) 95/100])
    (incidence) <- (incidence-from-habits newHabits)
    return (incidence))
\end{lstlisting}
  \end{minipage}
  \caption{Smoking example query.\label{fig:smoking-example}}
\end{figure}

Internally, the code computing incidence gets rewritten as in \Cref{fig:rewrite-smoking}.

\begin{figure}[ht]
    \begin{minipage}{0.8\textwidth}
    \begin{lstlisting}[language=Racket]
(define (incidence-from-habits habits)
  (do 
    (tar) <- (do 
        (smoking1 tar2 cancer3) <- survey
        () <- (observe habits smoking1)
        return (tar2))
    (cancer) <- (do 
        (cancer11 smoking12) <- (do 
            (smoking) <- (do 
                (smoking5 tar6 cancer7) <- survey
                return (smoking5))
            (cancer) <- (do 
                (smoking8 tar9 cancer10) <- survey
                () <- (observe tar tar9)
                () <- (observe smoking smoking8)
                return (cancer10))
            return (cancer smoking))
        return (cancer11))
    return (cancer)))\end{lstlisting} 
  \end{minipage}
  \caption{Automatic rewriting of part of \Cref{fig:smoking-example}.\label{fig:rewrite-smoking}}
\end{figure}

\clearpage
\subsection{Nested causal effects}

Our final example corresponds to Pearl's \emph{napkin problem}, which is not usually provided with a real-world interpretation. Its purpose is to show that the rewriting algorithm does handle nested cases and translates them to probabilistic programming primitives.

\begin{figure}[ht]
  \begin{minipage}{0.8\textwidth}
    \begin{lstlisting}[language=Racket]
(intervene p
  withModel (do
                u1 <- ()
                u2 <- ()
                w <- (u1 u2)
                z <- (w)
                x <- (z u1)
                y <- (x u2)
                return (w z x y))
  setting (x) to ('i) in (y))
\end{lstlisting}
\end{minipage}
\caption{``Napkin'' problem example.\label{fig:napkin-example}}
\end{figure}

\begin{figure}[ht]
  \begin{minipage}{0.8\textwidth}
    \begin{lstlisting}[language=Racket]
(do 
  (x16 y17) <- (do 
      (x13 w14 y15) <- (do 
          (w) <- (do 
              (w1 z2 x3 y4) <- p
              return (w1))
          (x) <- (do 
              (w5 z6 x7 y8) <- p
              () <- (observe z z6)
              () <- (observe w w5)
              return (x7))
          (y) <- (do 
              (w9 z10 x11 y12) <- p
              () <- (observe x x11)
              () <- (observe z z10)
              () <- (observe w w9)
              return (y12))
          return (x w y))
      return (x13 y15))
  () <- (observe 'i x16)
  return (y17))
\end{lstlisting}
  \end{minipage}
  \caption{Automatic rewriting of \Cref{fig:napkin-example}.\label{fig:rewriting-napkin}}
\end{figure}

\end{toappendix}

\section{Related work}

\emph{Causal programming} has been the subject of recent theoretical work \cite{wangtyped} and implementations \cite{chirho2023software,tavares2021language}, but literature towards a denotational semantics of causal probabilistic programming had not yet explicitly addressed the problem of identifiability.

We introduce a simple syntax and algebraic denotational semantics for causal programming. Our technique is not to consider \emph{causal programming} as a new paradigm requiring new primitives, but to introduce derived constructs on top of non-associative probabilistic programming with normalized semantics \cite{dilavore2026normalized}.

\section*{Acknowledgements}

Parts of this article build upon previous joint work with Márk Széles and Elena Di Lavore \cite{dilavore2026normalized}. The author thanks Márk Széles, Elena Di Lavore, Wessel de Weijer, and Théo Wang for many helpful comments and discussion.

\subsection*{Funding.} Mario Román was supported by the Estonian Research Council grant PRG 3215 (``Post Cartesian Programming - Logic, Probability and Computation with String Diagrams''), and by the Advanced Research + Invention Agency (ARIA) Safeguarded AI Programme.

\bibliographystyle{halpha}
\bibliography{main.bib}

\end{document}